\documentstyle[12pt]{article}

\begin{document}
\begin{titlepage}

\title{The gravitational energy-momentum flux}

\author{J. W. Maluf$\,^{*}$ and F. F. Faria\\
Instituto de F\'{\i}sica, \\
Universidade de Bras\'{\i}lia\\
C. P. 04385 \\
70.919-970 Bras\'{\i}lia DF, Brazil\\
and\\
K. H. Castello-Branco\\
Instituto de F\'{\i}sica, Universidade de S\~ao Paulo\\
Caixa Postal 66318, 05315-970 S\~ao Paulo SP, Brazil\\}
\date{}
\maketitle

\begin{abstract}

We present a continuity equation for the gravitational
energy-momentum, which is obtained in the framework of the
teleparallel equivalent of general relativity. From this 
equation it follows a general definition
for the gravitational energy-momentum flux. This definition
is investigated in the context of plane waves and of cylindrical
Einstein-Rosen waves. We obtain the well known value for the
energy flux of plane gravitational waves, and conclude that the
latter exhibit features similar to plane electromagnetic waves.

\end{abstract}
\thispagestyle{empty}
\vfill
\par
\bigskip
\noindent (*) e-mail: wadih@fis.unb.br\par
\end{titlepage}
\newpage

\noindent

\section{Introduction}
\bigskip
Attempts to define the gravitational energy in tetrad theories of
gravity were first put forward by M\o ller \cite{Mol}, who noticed
that the tetrad description of the gravitational field allows a more
satisfactory treatment of the gravitational energy-momentum. He
observed that a suitable expression constructed out of the tetrad
field and of the torsion tensor could possibly yield a definition
for the gravitational energy density. The torsion tensor cannot be
made to vanish at a point by a coordinate transformation. This
fact refutes the usual argument against the nonlocalizability of the
gravitational energy, and which rests on the reduction of the metric
tensor to the Minkowski metric tensor at a point in space-time by
means of a coordinate transformation.

It is well known that the teleparallel equivalent of general
relativity (TEGR)
\cite{Hay,Hehl,Kop,Muller,Nester1,Nester2,Maluf1,Per,MB,YO}
provides an alternative description of Einstein's general
relativity, in which the gravitational field is described by the
tetrad field. In fact the very first attempt to construct a
theory of the gravitational field in terms of a set
of four linearly independent vector fields in the Weitzenb\"ock
geometry \cite{Weit} is due to Einstein \cite{Einstein}.

A well posed and mathematically consistent expression for the
gravitational energy has been investigated \cite{Maluf2}. It arises
in the realm of the Hamiltonian formulation of the TEGR \cite{Maluf3}
and satisfies several crucial requirements for any acceptable
definition of gravitational energy. The definition of the
gravitational energy-momentum is obtained from the constraint
equations of the theory. The results so far obtained
indicate that such expression is physically relevant. 

In any small neighborhood of space the gravitational field
can be considered constant and uniform. The principle of
equivalence asserts that in such neighborhood it is always possible
to choose a reference frame in which the gravitational effects  
are not observed (we are adopting Einstein's version of the
principle of equivalence \cite{Norton}).
Thus in such reference frame we should not
detect any form of gravitational energy. Therefore it is reasonable
to expect that the localizability of the gravitational energy depends
on the reference frame, but not on the coordinate system. In fact any
other form of relativistic energy depends on the reference frame.
It turns out that the gravitational energy definition presented in
Ref. \cite{Maluf2} displays the feature discussed above, namely, it
depends on the reference frame. More precisely,
it depends on the choice of a global set of tetrad fields since the
energy expression is not invariant under local SO(3,1)
transformations of the tetrad field, but is invariant
under coordinate transformations of the three-dimensional spacelike
hypersurface (reference frames are better conceived in terms of
fields of vector bases \cite{Aldrovandi}).

It is pointed out in Ref. \cite{Maluf2} the
existence of a preferred global frame that allows a proper
discussion of the energy, momentum and angular momentum of the
gravitational field, and that  will be briefly
recalled in Sec. II of this article. In this section we will also
present an alternative interpretation of the tetrad field, based
on the relation between inertial and noninertial frames. In
Sec. III we consider the Lagrangian field equations of the TEGR and
obtain the continuity equation for the gravitational
energy-momentum. From
this equation it follows a definition for the gravitational
energy-momentum flux. The latter is applied in Secs. IV and V to
plane waves and Einstein-Rosen waves, respectively. By means of
simple calculations we obtain the well known value of the
gravitational energy flux carried by plane gravitational waves,
and we show that this value coincides with the flux of gravitational
momentum along the propagation direction, in similarity with
plane electromagnetic waves.

Notation: space-time indices $\mu, \nu, ...$ and SO(3,1)
indices $a, b, ...$ run from 0 to 3. Time and space indices are
indicated according to
$\mu=0,i,\;\;a=(0),(i)$. The tetrad field $e^a\,_\mu$ 
yields the definition of the torsion tensor:  
$T^a\,_{\mu \nu}=\partial_\mu e^a\,_\nu-\partial_\nu e^a\,_\mu$.
The flat, Minkowski space-time  metric is fixed by
$\eta_{ab}=e_{a\mu} e_{b\nu}g^{\mu\nu}= (-+++)$.        

\bigskip

\section{An interpretation of the tetrad field}

For a given space-time metric tensor $g_{\mu\nu}$ the
tetrad field is defined by the relation

\begin{equation}
e^a\,_\mu e^b\,_\nu \eta_{ab}\,=\,g_{\mu\nu}\;,
\label{1}
\end{equation}

\noindent where $\eta_{ab}$ is the
flat Minkowski metric tensor. For a metric tensor
$g_{\mu\nu}$ there exists an infinit set of tetrad fields that
satisfy the relation above, and which are related by a local Lorentz
transformation, $\tilde e^a\,_\mu(x)=\Lambda^a\,_b(x) e^b\,_\mu(x)$.
At every space-time point the SO(3,1) matrices $\Lambda^a\,_b$
satisfy

\begin{equation}
\Lambda^a\,_c \Lambda^b\,_d\, \eta_{ab}\,=\,\eta_{cd}\;.
\label{2}
\end{equation}
Tetrad fields are necessary in order to establish the coupling of
Dirac spinor fields with the gravitational field
\cite{Weyl,Schwinger}.

Let us consider the construction of tetrad fields for the flat
Minkowski space-time. For the latter we have the relation

\begin{equation}
e^a\,_\mu e^b\,_\nu\eta_{ab}=\eta_{\mu\nu}\;.
\label{3}
\end{equation}

We note that  Eq. (3) for the tetrad field is similar to Eq. (2)
for the matrices $\Lambda^a\,_b$. Therefore the tetrad field in flat
space-time can be considered as a transformation from an inertial
frame with coordinates $x^\mu$ to another inertial frame with
coordinates $q^a$ such that $dq^a=e^a\,_\mu dx^\mu$ and
$ e^a\,_\mu=\partial_\mu q^a$. In this case it is formally possible
to integrate $dq^a=e^a\,_\mu dx^\mu$ in order to obtain
$q^a=q^a(x^\mu)$ everywhere. Because of this property the
trasformation $x^\mu \rightarrow q^a$ is called holonomic. 

In the TEGR \cite{Maluf2}
every manifestation of the gravitational field other than the flat
Minkowski space-time is characterized by a set of tetrad fields
such that $\partial_\mu e^a\, _\nu - \partial_\nu e^a\,_\mu \ne 0$,
in  which case the tetrad field is no longer given by gradient
functions of the type $ e^a_\mu=\partial_\mu q^a$. In this case the
transformation $dq^a=e^a\,_\mu dx^\mu$ cannot be globally integrated,
and the transformation is called anholonomic. Therefore a nontrivial
manifestation of the gravitational field is characterized by a
nonvanishing torsion tensor,
$T^a\,_{\mu \nu}=\partial_\mu e^a\,_\nu-\partial_\nu e^a\,_\mu \ne 0$
\cite{Maluf2}. The four-dimensional space with coordinate
differentials $dq^a$,
endowed with Minkowski's metric tensor, will be denoted the
reference space-time.

Let the tetrad field be such that
$T^a\,_{\mu \nu} \ne 0$.
In any sufficiently small neighborhood of the space-time the
transformation $\Delta x^\mu \simeq e_a\,^\mu \Delta q^a$, where
$e_a\,^\mu$ are the inverses of $e^a\,_\mu$, may be
interpreted as a transformation from a local inertial frame with
coordinate differentials $\Delta q^a$ to a noninertial one with
curvilinear coordinate differentials $\Delta x^\mu$. Therefore in
such small neighborhood of the space-time the tetrad
field measures the deviation of the physical space-time from a
hypothetical flat space-time. From this point of view,
the principle of equivalence is naturally built into the kinematical
description of the space-time, since it establishes the local
equivalence of accelerated, noninertial frames with the
gravitational field.

We consider again the flat Minkowski space-time. In Sec. IV of
Ref. \cite{Maluf2} it is discussed the six conditions on the tetrad
field such that the reference space-time with coordinates $q^a$ is
neither related by a boost transformation nor rotating with respect
to the physical space-time with coordinates $x^\mu$. These 
conditions uniquely fix the tetrad field. In cartesian coordinates
they are given by

\begin{equation}
e_{(i)j}=e_{(j)i}\,,
\label{4}
\end{equation}

\begin{equation}
e_{(i)}\,^0=0\,.
\label{5}
\end{equation}

Let us consider a transformation between two cartesian space-time
coordinates $x^\mu$ and $q^a=q^a(x^\mu)$ such that $q^a$ is
rotating with respect to $x^\mu$. The tetrad field
$e^a\,_\mu=\partial_\mu q^a$ will acquire off diagonal components
in the spatial sector $\lbrace e_{(i)j} \rbrace$ \cite{Maluf2}.
Therefore the
imposition of Eq. (4) prevents the rotation between the two
cartesian space-times. Another transformation of general character
between $q^a$ and $x^\mu$ is a Lorentz boost. It is easy to see
that the imposition of the time gauge condition
$e_{(i)}\,^0=0=e^{(0)}\,_k$ prevents the two space-times from
beeing related by a boost transformation \cite{Maluf2}.

Conditions (4) and (5) will be assumed to hold also for an arbitrary
space-time metric tensor. They have proved to be essential in the
evaluation of the irreducible mass of the Kerr black
hole \cite{Maluf2}. Therefore for a given metric tensor $g_{\mu\nu}$
they establish a preferred global frame that allows a suitable
discussion of the energy, momentum and angular momentum of the
gravitational field.

\bigskip
\section{The continuity equation and the gravitational
energy-momentum flux}

The vacuum Lagrangian density of the TEGR  is given by

\begin{eqnarray}
L(e_{a\mu})&=& -k\,e\,({1\over 4}T^{abc}T_{abc}+
{1\over 2} T^{abc}T_{bac} -T^aT_a) \nonumber \\
&\equiv&-k\,e \Sigma^{abc}T_{abc}\;,
\label{6}
\end{eqnarray}
where $k=1/(16 \pi)$, $\Sigma^{abc}$ is defined by

\begin{equation}
\Sigma^{abc}={1\over 4} (T^{abc}+T^{bac}-T^{cab})
+{1\over 2}( \eta^{ac}T^b-\eta^{ab}T^c)\;,
\label{7}
\end{equation}
and $T^a=T^b\,_b\,^a$. The quadratic combination
$\Sigma^{abc}T_{abc}$ is proportional to the scalar curvature
$R(e)$, except for a total divergence \cite{Maluf1}.

The vacuum field equations read

\begin{equation}
{{\delta L}\over {\delta e^{a\mu}}}\;=\;
e_{a\lambda}e_{b\mu}\partial_\nu(e\Sigma^{b\lambda \nu})-
e(\Sigma^{b \nu}\,_aT_{b\nu \mu}-
{1\over 4}e_{a\mu}T_{bcd}\Sigma^{bcd})
\;=\;0\;.
\label{8}
\end{equation}
It is possible to prove by explicit calculations that

\begin{equation}
{{\delta L}\over {\delta e^{a\mu}}}\; \equiv \;{1\over 2}\,e\,
\biggl\{ R_{a\mu}(e)-{1\over 2}e_{a\mu}R(e)\biggr\}\;,
\label{9}
\end{equation}
The quantities on the right hand side of Eq. (9) are constructed
out of the curvature tensor $R_{ab\mu\nu}(e)$.

The gravitational energy-momentum is obtained in the framework of
the Hamiltonian formulation of the TEGR \cite{Maluf3}. The
constraint equations of the theory are interpreted as equations that
define the energy, momentum and angular momentum of the
gravitational field. The gravitational energy-momentum is defined by
\cite{Maluf2}

\begin{equation}
P^a=-\int_V d^3x\,\partial_j \Pi^{aj}\;,
\label{10}
\end{equation}

\noindent where $\Pi^{ai}$ is the momentum canonically conjugated
to $e_{ai}$ and reads

\begin{eqnarray}
\Pi^{ak}&=&-4ke\Sigma^{a0k} \nonumber \\
&=& k\,e\biggl\{
g^{00}(-g^{kj}T^a\,_{0j}-
e^{aj}T^k\,_{0j}+2e^{ak}T^j\,_{0j}) \nonumber \\
&+&g^{0k}(g^{0j}T^a\,_{0j}+e^{aj}T^0\,_{0j})
\,+e^{a0}(g^{0j}T^k\,_{0j}+g^{kj}T^0\,_{0j}) \nonumber \\
&-& 2(e^{a0}g^{0k}T^j\,_{0j}+e^{ak}g^{0j}T^0\,_{0j})
-g^{0i}g^{kj}T^a\,_{ij} \nonumber \\
&+&e^{ai}(g^{0j}T^k\,_{ij}-
g^{kj}T^0\,_{ij})-2(g^{0i}e^{ak}-g^{ik}e^{a0})
T^j\,_{ji} \biggr\}\;.
\label{11}
\end{eqnarray}
Equation (10) satisfies all requirements that are expected from
any definition of gravitational energy. This definition:
(i) vanishes for the Minkowski space-time, in which case the
tetrad field is given by gradient functions of the type
$e^a\,_\mu=\partial_\mu q^a$; (ii) yields the ADM
(Arnowitt-Deser-Misner) and Bondi energy in the appropriate limits
\cite{Maluf4}; (iii) yields the appropriate value for weak and
spherically symmetric gravitational fields; (iv) yields the
irreducible mass of the Kerr black hole \cite{Maluf2}; (v)
is invariant under coordinate transformations on the
three-dimensional spacelike hypersurface, and transforms as a
vector under the global SO(3,1) group.

We proceed now to derive the continuity equation for the
gravitational energy-momentum. The field equations (8) can be
rewritten by multiplying it by the inverse tetrad fields. We
obtain

\begin{equation}
\partial_\nu(-4ke\Sigma^{a\lambda \nu})=
-ke\,e^{a\mu}(4\Sigma^{b\nu\lambda}T_{b\nu\mu}-
\delta^\lambda_\mu \Sigma^{bcd}T_{bcd})\;.
\label{12}
\end{equation}

\noindent Restricting $\lambda$ to spatial components, i.e.,
$\lambda=j$, we find

\begin{equation}
-\partial_0(-4ke\Sigma^{a0j})-\partial_k(-4ke\Sigma^{akj})
=-kee^{a\mu}(4\Sigma^{bcj}T_{bc\mu}-
\delta^j_\mu \Sigma^{bcd}T_{bcd})\;.
\label{13}
\end{equation}

\noindent Note that $\Sigma^{abc}=-\Sigma^{acb}$.
Taking the divergence of the equation above with
respect to $j$ yields

\begin{equation}
-\partial_j \partial_0(-4ke\Sigma^{a0j})=
-k\partial_j\lbrack e e^{a\mu}
(4\Sigma^{bcj}T_{bc\mu}-\delta^j_\mu \Sigma^{bcd}T_{bcd})\rbrack\;,
\label{14}
\end{equation}

\noindent or

\begin{equation}
-\partial_0(\partial_j\Pi^{aj})=
-k\partial_j\lbrack e e^{a\mu}
(4\Sigma^{bcj}T_{bc\mu}-\delta^j_\mu \Sigma^{bcd}T_{bcd})\rbrack\;.
\label{15}
\end{equation}

\noindent By integrating Eq. (15) in a space volume V we arrive at

\begin{equation}
{d \over {dt}}\biggl[ -\int_V d^3x\,\partial_j \Pi^{aj}\biggr]=
-k\int_S dS_j [ ee^{a\mu}(4\Sigma^{bcj}T_{bc\mu}-
\delta^j_\mu \Sigma^{bcd}T_{bcd}) ]\;.
\label{16}
\end{equation}

\noindent The time derivative of the gravitational energy in a
volume $V$ of the three-dimensional spacelike hypersurface is
define to be minus the gravitational energy-momentum flux
$\Phi^{a}$,

\begin{equation}
{d \over {dt}}\biggl[ -\int_V d^3x\,\partial_j \Pi^{aj}\biggr]=
-\Phi^a\;,
\label{17}
\end{equation}
where

\begin{equation}
\Phi^a=k\int_S dS_j[ ee^{a\mu}(4\Sigma^{bcj}T_{bc\mu}-
\delta^j_\mu \Sigma^{bcd}T_{bcd}) ]\;.
\label{18}
\end{equation}

We will give one step further and assume that Eq. (18) is defined
also for open surfaces $S$. For later purposes we define the
gravitational energy-momentum flux density $\phi^{aj}$,

\begin{equation}
\phi^{aj}=k [ ee^{a\mu}(4\Sigma^{bcj}T_{bc\mu}-
\delta^j_\mu \Sigma^{bcd}T_{bcd}) ]\;,
\label{19}
\end{equation}
which represents the $a$ component of the flux density in the
$j$ direction.

In Secs. IV and V we will carry out concrete applications
of Eqs. (18) and (19). Here we just mention that for tetrad fields
with appropriate boundary conditions, in an asymptotically flat
space-time, Eq. (16) leads to the total conservation of the
gravitational energy-momentum.

Before closing this section we briefly
mention that Eq. (19) is similar 
to the gravitational gauge current defined in Ref. \cite{Per2},
but the interpretations of $\phi^{aj}$ are different in the two
situations. In the latter Ref. the quantity similar to Eq. (19)
is taken to represent the energy and
momentum of the gravitational field, whereas in the present
context it really represents the flux of energy and momentum.

\bigskip
\section{The energy-momentum flux of linear plane
gravitational waves}

We will consider initially the simplest realization of a linear
plane wave, that is a solution of Einstein's equations. The energy
carried away by these waves is known in the literature
\cite{Schutz}. As in the latter Ref., we will restrict
considerations to just one polarization of the wave, which is
described by the metric tensor

\begin{equation}
ds^2=-dt^2+\lbrack 1+f_+(t-z)\rbrack dx^2+
\lbrack 1-f_+(t-z)\rbrack dy^2+dz^2\,.
\label{20}
\end{equation}
where $(f_+)^2<<1$. 
The tetrad field that satisfies conditions (4) and (5) and that
yields the metric tensor above reads (lines are labelled by $a$
and columns by $\mu$)

\begin{equation}
e^a\,_\mu(t,x,y,z)=\pmatrix{1&0&0&0\cr
0&\sqrt{1+f_+}&0&0\cr
0&0&\sqrt{1-f_+}&0\cr
0&0&0&1\cr}\;.
\label{21}
\end{equation}

\noindent The determinant $e=\det(e^a\,_\mu)$ is given by
$e =\sqrt{1+f_+}\sqrt{1-f_+}$. The contravariant
components of the metric tensor (which is given in diagonal form)
are  $g^{00}= -1$, $g^{11}=1/(1+f_+)$,
$g^{22}=1/(1-f_+)$, $g^{33}=1$.

The nonvanishing components of the torsion tensor are easily
calculated,

$$T_{(1)01}= {{\dot f}\over {2\sqrt{1+f_+}}},$$

$$T_{(2)02}=-{{\dot f}\over {2\sqrt{1-f_+}}},$$

$$T_{(1)13}=-{{f_+^\prime} \over {2\sqrt{1+f_+}}},$$

$$T_{(2)23}= {{f_+^\prime} \over {2\sqrt{1-f_+}}},$$

\noindent where the dot and the prime denote derivatives with
respect to the coordinates $t$ and $z$, respectively.
From the components above we evaluate
$T_{\lambda\mu\nu}=e^a\,_\lambda T_{a\mu\nu}$. We find

\begin{equation}
T_{101}={1\over 2}{\dot f_+}\;,
\label{22}
\end{equation}

\begin{equation}
T_{202}=-{1\over 2} {\dot f_+}\;,
\label{23}
\end{equation}

\begin{equation}
T_{113}=-{1\over 2} f_+^\prime\;,
\label{24}
\end{equation}

\begin{equation}
T_{223}={1\over 2} f_+^\prime\;.
\label{25}
\end{equation}

We first calculate the energy flux $\Phi^{(0)}$. It is given by

\begin{eqnarray}
\Phi^{(0)}&=&k\int_S dS_j[ ee^{{(0)}\mu}(4\Sigma^{bcj}T_{bc\mu}-
\delta^j_\mu \Sigma^{bcd}T_{bcd}) ] \nonumber \\
&=&4k\int_S dS_j\,e e^{(0)0}{\delta^\mu_0} \Sigma^{\lambda \nu j}
T_{\lambda \nu \mu} \nonumber \\
&=&-4k\int_S dS_j\, e g^{00} e^{(0)}\,_0
(\Sigma^{11j}T_{101}+\Sigma^{22j}T_{202})\;.
\label{26}
\end{eqnarray}
In order to evaluate the tensor $\Sigma^{\mu\nu\lambda}$ we need
the traces
$T^\mu=g^{\mu\nu}T_\nu=g^{\mu\nu}T^\lambda\,_{\lambda \nu}$.
We find

\begin{equation}
T^1=T^2=0\;,\;\;\;\;\;\;
T^3=g^{11}g^{33}T_{113}+g^{22}g^{33}T_{223}
\label{27}
\end{equation}
By considering $j=1$ in Eq. (26) we obtain

$$\Sigma^{221}T_{202}=T_{202}\biggl[
{1\over 4}(T^{221}+T^{221}-T^{122})+
{1\over 2}(g^{21}T^2-g^{22}T^1)\biggr]=0\;,$$
because $T^{221}=g^{22}g^{22}g^{11}T_{221}=0$, etc.
Similarly, for $j=2$ in Eq. (26) we have

$$\Sigma^{112}T_{101}=
T_{101}\biggl[{1\over 4}(T^{112}+T^{112}-T^{211})+
{1\over 2}(g^{12}T^1-g^{11}T^2)\biggr]=0\;.$$
Therefore the only contribution to $\Phi^{(0)}$ comes from the
integration on $z$ = constant surfaces, namely, on the surface
orthogonal to the propagation of the wave. We have

\begin{eqnarray}
\Phi^{(0)}&=&-4k\int_SdS_3\,e\, g^{00} e^{(0)}\,_0
(\Sigma^{113}T_{101}+\Sigma^{223}T_{202})\nonumber \\
&=&4k\int_SdS_3\,e\biggl\{
T_{101}\biggl[{1\over 4}(T^{113}+T^{113}-T^{311})+
{1\over 2}(g^{13}T^1-g^{11}T^3)\biggr]  \nonumber \\
&+&T_{202}\biggl[{1\over 4}(T^{223}+T^{223}-T^{322})+
{1\over 2}(g^{23}T^2-g^{22}T^3)\biggr] \biggr\} \nonumber \\
&=&-2k\int_S dS_3\,e\biggl[g^{11}g^{22}g^{33}(
T_{101}T_{223}+T_{202}T_{113})\biggr]\;.
\label{28}
\end{eqnarray}
By substituting  Eqs. (22-25) into Eq. (28) we find

\begin{equation}
\Phi^{(0)}=-k\int_SdS_3\,{{\dot f_+ f_+^\prime}\over
\sqrt{1-(f_+)^2}}\;.
\label{29}
\end{equation}
Since the function $f_+$ is such that $f_+^2<<1$, the equation above
can be rewritten to a very good approximation as

\begin{equation}
\Phi^{(0)}\approx -k\int_SdS_3\, \dot f_+ f_+^\prime\;.
\label{30}
\end{equation}

Assuming now that the function $f_+$ can be written in terms of
an amplitude $A$ and a frequency $\omega$ as
\cite{Schutz}

\begin{equation}
f_+(t-z)=A\,\cos\omega(z-t)\;,
\label{31}
\end{equation}
we have

$$\dot f_+= A\omega\, \sin\omega(z-t)\;,\;\;\;\;\;\;\;\;
f_+^\prime= -A\omega\,\sin\omega(z-t)\;,$$
from what follows

\begin{equation}
\Phi^{(0)}={{A^2\omega^2}\over {16\pi}}
\sin^2\omega(z-t)\int_SdS_3\;,
\label{32}
\end{equation}
where $S$ is an arbitrary
$z$ = constant surface, and we have substituted
$k=1/(16\pi)$. We take the avarage value
$<\Phi^{(0)}>$ over a period $T$ by considering the integral

$$\int_0^T\,dt\, \sin^2(\omega t- \omega z)={T\over 2}\;.$$
Taking into account the integral above in Eq. (32) we obtain the
average flux density $<\phi^{(0)3}>$ per unit period $T$, flowing
along the $z$ direction,

\begin{equation}
{{<\phi^{(0)3}>}\over T}={{A^2\omega^2}\over {32\pi}}\;.
\label{33}
\end{equation}
This is precisely the value obtained in the literature
by means of a completely different analysis of the
the energy flux of plane, linearised
gravitational waves \cite{Schutz}.

By evaluating the momentum flux
components $\Phi^{(i)}$ we arrive at a quite
interesting result. For this purpose we first need to calculate
the expression of the scalar $\Sigma^{bcd}T_{bcd}$, which is
easily obtained as

\begin{equation}
\Sigma^{bcd}T_{bcd}=-2(g^{11}g^{22}g^{33}T_{113}T_{223}+
g^{00}g^{11}g^{22}T_{101}T_{202})\,.
\label{34}
\end{equation}

The momentum flux $\Phi^{(1)}$ is given by

\begin{eqnarray}
\Phi^{(1)}&=&k\int_S dS_j\,\lbrack ee^{(1)1}(
4\Sigma^{\mu\nu j}T_{\mu\nu 1}
-\delta^j_1\Sigma^{bcd}T_{bcd})\rbrack \nonumber \\
&=& k\int_S dS_1\,\lbrack ee^{(1)1}(
4\Sigma^{\mu\nu 1}T_{\mu\nu 1}
-\Sigma^{bcd}T_{bcd})\rbrack \nonumber \\
&+&k\int_S dS_2 \lbrack 4e e^{(1)1}
\Sigma^{\mu\nu 2}T_{\mu\nu1}\rbrack 
+k\int_S dS_3 \lbrack 4e e^{(1)1}
\Sigma^{\mu\nu 3}T_{\mu\nu1}\rbrack\;.
\label{35}
\end{eqnarray}
After some simple calculations we obtain

\begin{equation}
\Sigma^{\mu\nu 1}T_{\mu\nu 1}=
-{1\over 2}(g^{00}g^{11}g^{22}T_{101}T_{202}+
g^{11}g^{22}g^{33}T_{113}T_{223})\;,
\label{36}
\end{equation}
and

\begin{equation}
\Sigma^{\mu\nu 2}T_{\mu\nu 1}=\Sigma^{\mu\nu 3}T_{\mu\nu 1}=0\,.
\label{37}
\end{equation}
By substituting Eqs. (34), (36) and (37) into Eq. (35) we conclude
that $\Phi^{(1)}=0$.

A similar result is obtained for $\Phi^{(2)}$, which reads

\begin{eqnarray}
\Phi^{(2)}&=&k\int_S dS_j\,\lbrack ee^{(2)2}(
4\Sigma^{\mu\nu j}T_{\mu\nu 2}
-\delta^j_2\Sigma^{bcd}T_{bcd})\rbrack \nonumber \\
&=& k\int_S dS_1\,\lbrack 4 ee^{(2)2}
\Sigma^{\mu\nu 1}T_{\mu\nu 2}\rbrack
+k\int_S dS_2 \lbrack e e^{(2)2}(
4\Sigma^{\mu\nu 2}T_{\mu\nu2}
-\Sigma^{bcd}T_{bcd})\rbrack \nonumber \\
&+&k\int_S dS_3 \lbrack 4e e^{(2)2}
\Sigma^{\mu\nu 3}T_{\mu\nu2}\rbrack\;.
\label{38}
\end{eqnarray}
In this case we have

\begin{equation}
\Sigma^{\mu\nu 2}T_{\mu\nu 2}=
-{1\over 2}(g^{00}g^{11}g^{22}T_{101}T_{202}+
g^{11}g^{22}g^{33}T_{113}T_{223})\;,
\label{39}
\end{equation}

\begin{equation}
\Sigma^{\mu\nu 1}T_{\mu\nu 2}=\Sigma^{\mu\nu 3}T_{\mu\nu 2}=0\;.
\label{40}
\end{equation}
By considering Eqs. (39) and (40) we conclude that Eq. (38)
reduces to $\Phi^{(2)}=0$.

Finally, for $\Phi^{(3)}$ we have

\begin{eqnarray}
\Phi^{(3)}&=&k\int_S dS_j\,\lbrack ee^{(3)3}(
4\Sigma^{\mu\nu j}T_{\mu\nu 3}
-\delta^j_3\Sigma^{bcd}T_{bcd})\rbrack \nonumber \\
&=& k\int_S dS_1\,\lbrack 4 ee^{(3)3}
\Sigma^{\mu\nu 1}T_{\mu\nu 3}\rbrack
+k\int_S dS_2 \lbrack 4 e e^{(3)3}
\Sigma^{\mu\nu 2}T_{\mu\nu 3} \rbrack \nonumber \\
&+&k\int_S dS_3 \lbrack e e^{(3)3}(4
\Sigma^{\mu\nu 3}T_{\mu\nu 3} -\Sigma^{bcd}T_{bcd})\rbrack\;.
\label{41}
\end{eqnarray}
Simple calculations yield

\begin{equation}
\Sigma^{\mu\nu 1}T_{\mu\nu 3}=\Sigma^{\mu\nu 2}T_{\mu\nu 3}=0\;,
\label{42}
\end{equation}
and

\begin{equation}
\Sigma^{\mu\nu 3}T_{\mu\nu 3}=
-g^{11}g^{22}g^{33}T_{113}T_{223}\;.
\label{43}
\end{equation}
Substitution of Eqs. (34), (42) and (43) into Eq. (41) leads to

\begin{equation}
\Phi^{(3)}=k\int_S dS_3\lbrack e e^{(3)}\,_3 g^{33}(
2g^{00}g^{22}g^{22}T_{101}T_{202}-
2g^{11}g^{22}g^{33}T_{113}T_{223})\rbrack\;,
\label{44}
\end{equation}
With the help of Eqs. (22-25) we find

\begin{equation}
\Phi^{(3)}=k\int_S dS_3 {{(f_+^\prime)^2+(\dot f_+)^2}\over
{2\sqrt{1-(f_+)^2}}}\;,
\label{45}
\end{equation}
where $S$ is now an arbitrary $z$ = constant surface.
Taking into account Eq. (31) we observe that $(f_+^\prime)^2+
(\dot f_+)^2=(f_+^\prime + \dot f_+)^2-2f_+^\prime \dot f_+=
-2f_+^\prime \dot f_+$, since $f_+^\prime=-\dot f_+$.
Thus we find

\begin{equation}
\Phi^{(3)}=-k\int_SdS_3\,{{\dot f_+ f_+^\prime}\over
\sqrt{1-(f_+)^2}}\approx -k\int_SdS_3\, \dot f_+ f_+^\prime\;,
\label{46}
\end{equation}
in similarity to Eqs. (29) and (30).

Therefore the expression for $\Phi^{(3)}$ is identical to the one
for $\Phi^{(0)}$. We obtain

\begin{equation}
\Phi^{(3)}={{A^2\omega^2}\over {16\pi}}
\sin^2\omega(z-t)\int_SdS_3\;.
\label{47}
\end{equation}
Taking again the avarage value
$<\Phi^{(3)}>$ over a period $T$  we obtain average momentum
flux density $<\phi^{(3)3}>$ per unit period $T$, flowing along
the $z$ direction,

\begin{equation}
{{<\phi^{(3)3}>}\over T}={{A^2\omega^2}\over {32\pi}}\;.
\label{48}
\end{equation}

We conclude that the energy and momentum flux density per unit
time of a linear plane wave, along the direction of propagation,
are the same. It follows that the fluxes satisfy the relation

\begin{equation}
\Phi^a \Phi^b \eta_{ab}=0\;.
\label{49}
\end{equation}
We note that a similar relation must be satisfied by the
energy-momentum four-vector of massless particles. We also remark
that plane electromagnetic waves display the same feature, namely,
the fluxes of energy and momentum are the same, in natural
units (c=1).

\bigskip
\section{The energy flux of Einstein-Rosen waves}

Einstein-Rosen waves arise as exact solutions of Einstein's field
equations \cite{ER}. They describe cylindrical waves determined by
two functions $\gamma(\rho, t)$ and $\psi(\rho, t)$. In cylindrical
coordinates the waves are described by

\begin{equation}
ds^2=e^{2(\gamma-\psi)}(-dt^2+d\rho^2)+\rho^2e^{-2\psi}d\phi^2+
e^{2\psi}dz^2\;.
\label{50}
\end{equation}
The functions $\gamma$ and $\psi$ satisfy

\begin{equation}
\psi^{\prime\prime} +{1\over \rho}\psi^\prime -\ddot{\psi}=0\;,
\label{51}
\end{equation}

\begin{equation}
\gamma^\prime=\rho\lbrack (\psi^\prime)^2+(\dot \psi)^2\rbrack\;,
\label{52}
\end{equation}

\begin{equation}
\dot\gamma=2\rho\,\psi^\prime \dot \psi\;,
\label{53}
\end{equation}
where now the prime denotes differentiation with respect to $\rho$.

The expressions for the gravitational energy contained within
a cylindrical region of arbitrary length $L$ and radius $\rho$,
around the z axis, and the corresponding energy flux are very
simple, as we will see. The procedure for obtaining these
quantities is also very simple. First we determine the tetrad field
that satisfies conditions (4) and (5), and that leads to Eq. (50).
It is given by

\begin{equation}
e^a\,_\mu(t,\rho,\phi,z)=\pmatrix{A&0&0&0\cr
0&A\cos\phi&-\rho C\sin\phi&0\cr
0&A\sin\phi&\rho C \cos\phi&0\cr
0&0&0&B\cr}\,,
\label{54}
\end{equation}
where

\begin{equation}
A=e^{\gamma-\psi}\,,\;\;\;\;\;\;B=e^\psi\,,\;\;\;\;\;\;
C=e^{-\psi}\;.
\label{55}
\end{equation}
It follows that $e=\det(e^a\,_\mu)=\rho e^{2(\gamma - \psi)}$. The
nonvanishing components of the torsion tensor are

$$T_{(0)01}=A^\prime\;,$$

$$T_{(1)01}=\dot A \cos\phi\,,\;\;\;\;\;
T_{(1)02}=-\rho \dot C \sin\phi\,,\;\;\;\;\;
T_{(1)12}=(A-C-\rho C^\prime)\sin\phi\;,$$

$$T_{(2)01}=\dot A \sin\phi\,,\;\;\;\;\;
T_{(2)02}= \rho \dot C \cos\phi\,,\;\;\;\;\;
T_{(2)12}= -(A-C-\rho C^\prime) \cos\phi\,,$$

$$T_{(3)03}=\dot B\,,\,\,\,\,\,\,T_{(3)13}=B^\prime\;,$$
which in turn lead to

$$T_{001}=AA^\prime,\;\;\;\;\;T_{101}=A\dot A,\;\;\;\;\;
T_{202}=\rho^2 C \dot C,\;\;\;\;\;$$

$$T_{212}=\rho C(C+\rho C^\prime -A),\;\;\;\;\;
T_{303}=B\dot B,\;\;\;\;\;T_{313}=BB^\prime\,.$$

In order to calculate the gravitational energy associated with
the metric tensor (50) we can either calculate the $a=(0)$
component of Eq. (10) or, more simply, take advantage of an
equality that holds in the time gauge condition \cite{Maluf2},

\begin{equation}
-\int_V d^3x\,\partial_i \Pi^{(0)i}=
{1\over {8\pi}}\int_V d^3x\,\partial_i(eT^i)\,.
\label{56}
\end{equation}
The field quantities on the left hand side of the equation above
are defined on the space-time, and those on the right hand side are
defined on a three-dimensional spacelike hypersurface $\Sigma$.
We note that the tetrad field (54) satisfies the time
gauge condition because $e^{(0)}\,_k=0$. In view of
Gauss theorem we define the energy contained within a surface $S$
by \cite{Maluf5}

\begin{equation}
E={1\over {8\pi}}\int_S dS_i (eT^i)\,,
\label{57}
\end{equation}
where $e$ and $T^i$ are calculated on $\Sigma$.
Therefore we consider triads
$e_{(i)j}$ on a $t$ = constant surface,

\begin{equation}
e^{(i)}\,_j(t,\rho,\phi,z)=\pmatrix{
&A\cos\phi&-\rho C\sin\phi&0\cr
&A\sin\phi&\rho C \cos\phi&0\cr
&0&0&B\cr}\,,
\label{58}
\end{equation}

\noindent and construct $g_{ij}$ and the
inverses $g^{ij}$ and $e^{(i)j}$ on the spacelike hypersurface
$\Sigma$ (because the metric is diagonal, $g^{ij}$ on $\Sigma$
and on the space-time are the same).
In this case the determinant $e=\det(e_{(i)j})$ is given by
$e=\rho e^{\gamma -\psi}$, and the trace
$T^1=g^{11}T_1=g^{11}(g^{22}T_{221}+g^{33}T_{331})$ reads

\begin{equation}
T^1={1\over \rho}e^{-2(\gamma - \psi)}(e^\gamma -1)\;.
\label{59}
\end{equation}
We also have $T^2=T^3=0$. The gravitational energy enclosed by a
cylinder of length $L$ and arbitrary radius $\rho$ is obtained
by a trivial integration of Eq. (57),

\begin{equation}
E(t,\rho)={L\over 4} e^{-(\gamma-\psi)}(e^\gamma-1)\;.
\label{60}
\end{equation}

\noindent The energy per unit length $\varepsilon$ for very small
values of $\gamma$ and $\psi$ reads
$\varepsilon (t,\rho)\approx \gamma / 4$. In this
limiting case the latter value coincides with
the value obtained by Thorne \cite{Thorne} in the analysis of the
C-energy of Einstein-Rosen waves.

We proceed now to calculate the gravitational energy flux through
the cylindrical surface of length $L$ and radius $\rho$. In view of
Eq. (18) it is given by

\begin{eqnarray}
\Phi^{(0)}&=&k\int_S dS_1\,(4ee^{(0)0}
\Sigma^{bc1}T_{bc0})\nonumber \\
&=&-4k\int_S dS_1\,e e^{(0)0}(\Sigma^{\mu 21}T_{\mu 02}+
\Sigma^{\mu 31}T_{\mu 03}) \nonumber \\
&=&-4k\int_S dS_1\,e e^{(0)0}(\Sigma^{221}T_{202}
+\Sigma^{331}T_{303})\,.
\label{61}
\end{eqnarray}
In the evaluation of Eq. (61) we are taking into account 
definitions (54) and (55). After simple calculations we obtain

\begin{eqnarray}
\Sigma^{221}T_{202}+\Sigma^{331}T_{303}&=&
{1\over 2}g^{11}g^{22}g^{33}(T_{212}T_{303}
+T_{313}T_{202})\nonumber \\
&-&{1\over 2}(g^{00}g^{11}g^{22}T_{001}T_{202}
+g^{00}g^{11}g^{33}T_{001}T_{303})\;.
\label{62}
\end{eqnarray}

\noindent By substituting the expressions of $T_{\mu\nu\lambda}$
we eventually find

\begin{equation}
\Phi^{(0)}=-2k\int_Sd\phi dz\,\lbrack e^{-(\gamma - \psi)}
(e^{\gamma}-1) \dot \psi\rbrack\;.
\label{63}
\end{equation}
Considering a cylindrical surface of length $L$ and radius $\rho$
we obtain

\begin{equation}
\Phi^{(0)}=-{L\over 4}  e^{-(\gamma - \psi)}
(e^{\gamma}-1) \dot \psi\;.
\label{63}
\end{equation}

The simplicity of expressions (60) and (64) is an 
indication that the present framework is suitable for discussing
the energy-momentum properties of the gravitational field.

\bigskip

\section{Discussion}

We have derived a simple expression for the energy-momentum flux
of the gravitational field. This expression is obtained on the
assumption that Eq. (10) represents the energy-momentum of the
gravitational field on a volume $V$ of the three-dimensional
spacelike hypersurface. Of course the consistency of the present
results may be taken as a further, {\it a posteriori}
justification for Eq. (10) to represent the energy and momentum
of the gravitational field.

Application of Eq. (18) to
linear plane gravitational waves shows that the latter has
properties similar to plane electromagnetic waves: the energy
and momentum fluxes are the same, in natural units. This is
the main result of this article. Moreover,
the value of the energy flux obtained here is exactly the same
one derived in the literature by considering the energy supplied
by the waves to a nearly continuous distribution of oscillators
on a plane orthogonal to the direction of propagation of the
wave \cite{Schutz}. In the latter analysis it is evaluated the
reduction of the amplitude of the gravitational wave as it passes
through the configuration of oscillators describe above.
Einstein-Rosen waves were also analysed. The expressions of the
gravitational energy contained within a cylinder around the $z$
axis and the corresponding energy flux through the cylindrical
surface turned out to be simple as well. Altogether, the results
described above indicate the consistency of the present framework
in the investigation of the energy-momentum properties of the
gravitational field.

\bigskip
\bigskip
\noindent {\sl Acknowledgements}\par
\noindent K. H. C. B. and F. F. F. are grateful to the Brazilian
agencies FAPESP and CNPQ, respectively, for financial support.\par
\bigskip

\end{document}